# Thermophysical and magnetic properties of p- and n-type Ti-Ni-Sn based half-Heusler alloys

Md. Mofasser Mallik and Satish Vitta*

Department of Metallurgical Engineering and Materials Science

Indian Institute of Technology Bombay

Mumbai 400076; India.

**Abstract:**

A total of 5 different half-Heusler alloys, two p-type and two n-type with the fifth a charge compensated alloy have been designed and synthesized. The thermophysical properties of these alloys have been investigated in the range 10 K to 1000 K while the magnetic behavior has been studied up to 300 K. The electrical resistivity of all the alloys varies within the range 0.06 to 5 mΩ cm indicating that they are in the degenerate semiconductor limits. The temperature dependence of p-type alloys exhibits a transition from metallic to semiconducting behavior, typical of topological insulators. The transition is found to occur in the range 300 K to 500 K. The n-type and compensated alloys exhibit a weak metallic behavior in the complete temperature range. The Seebeck coefficient in the p-type alloys increases with temperature reaching a maximum value of 50 µV K$^{-1}$ while that of the n-type alloys increases continuously reaching a value of 45 µV K$^{-1}$ at ~ 800 K. The corresponding power factor of the n-type alloy reaches 900 µW m$^{-1}$ K$^{-2}$ at ~ 900 K compared to a maximum of ~ 250 µW m$^{-1}$ K$^{-2}$ at 700 K for the p-type alloy. Magnetically the p-type and n-type alloys are found to be paramagnetic while the compensated alloy exhibits a ferromagnetic behavior.

* Email for correspondence: **satish.vitta@iitb.ac.in**

**Introduction:**

Heusler alloys are an extremely interesting and highly versatile group of intermetallic compounds exhibiting a variety of electronic behaviors ranging from semiconducting to superconducting and ferromagnetic to non-magnetic. The electronic property versatility of these compounds is similar to that exhibited by perovskite oxide group of compounds which has a large variety of behaviors. A typical Heusler compound is a ternary alloy of transition metals belonging to groups 4-10, and a main group element of groups 13-15 in the ratio 1:1:1 or 1:2:1. Accordingly, they are termed as half-Heusler or full-Heusler alloys respectively and denoted as XYZ or $XY_2Z$ alloy. In this representation X is the most electropositive element while Z is the most electronegative element. Although different conventions are followed in the literature, in the present work this convention is followed as it does not lead to ambiguities, especially with respect to atomic position in the crystal structure. The half-Heusler alloys are known to be stable in a cubic crystal structure belonging to F-43m space group while the full-Heusler alloy has Fm-3m structure and in both cases the structure consists of 4 interpenetrating cubic sub-lattices. The structure has 3 unique positions for atoms to occupy: 4a (0,0,0), 4b(1/2,1/2,1/2) and 4c(1/4,1/4,1/4) and depending on which of the 3 atoms occupy these positions, 3 different structural variations are possible in these alloys. It should be noted here that since X and Z are most electropositive and electronegative elements respectively, the nature of interaction between these two elements will be ionic while other two interactions, XY and YZ will be predominantly covalent. Hence the electronic properties of these alloys are strong functions of positional occupancy of the different atomic species as well as the total valance electron count, VEC. The large possibility of substitutions with different VEC and positional occupancy opens an opportunity to tune the properties, including formation of topological insulators.(1-6) The half-Heusler alloys have an additional variable - presence of a vacant sub-lattice due to absence of an Y atom. The presence of these vacancies in the lattice facilitates reduction of thermal conductivity and makes them suitable candidates for thermoelectric energy conversion.(7-14) Apart from this application, the only half-Heusler alloy that has been found to exhibit a half-metallic ferromagnetic behavior is NiMnSb, excluding the rare-earth substituted alloys.(15-20) Hence the objective of the present work therefore has been to investigate both the thermoelectric related physical properties as well as study the feasibility of a ferromagnetic behavior in half-Heusler alloys without Mn or rare-earth elements in the compositions.

Among the several different half-Heusler alloys, n-type MNiSn and p-type MCoSb, where 'M' is a transitional metal belonging to group 4 elements have been extensively investigated for thermoelectric energy conversion.(21-26) Although these alloys exhibit relatively high thermoelectric figure of merit, zT, they have still not found large scale commercial application due to the variability in zT values as a result of processing related microstructural variations.(8, 23, 27-29) Also, they are two separate classes of materials with Sn and Sb as the two electronegative elements with Ni and Co as the Y transition metals respectively, which makes for a non-ideal thermoelectric couple. An ideal thermoelectric couple should have nearly identical composition for the two legs such that there is no significant mismatch in their mechanical, thermal fatigue or thermal expansion properties. Hence in the present work TiNiSn based half-Heusler alloys have been chosen as Zr and Hf containing alloys have been extensively investigated. The chemical composition of the alloys has been designed to exhibit either p-type or n-type or a compensated behavior by suitable substitution of not only Z element but also the X and Y elements. The criterion used for selecting the substitutional elements is rather simple in nature – all elements to the left of the main element with electron deficiency compared to the main element will impart a p-type character while elements to its right with an additional electrons will induce an n-type character. These substitutions should also result in reducing the overall thermal conductivity of the alloys due to alloy scattering, large atomic mass and size variations, in addition to vacancy related scattering.

The nature of charge carriers as well as their concentration have been predicted using TiNiSn as the base composition. The different chemical compositions that have been selected and their expected electrical transport behavior and magnetic moments are given in Table 1. According to this criterion, alloy P1 should exhibit a p-type behavior due to the valance electron difference between Ni and Co while alloy P2 should exhibit a p-type behavior due to both Sc and In substitution for X and Z atoms respectively. The charge carrier concentration in P2 should be greater than that in P1 alloy because of both X and Z substitutions in P2 compared to only Y substitution in P1. In the case of n-type alloys, N1 and N2, it is due to substitution of all the 3 elements Ti, Ni and Sn, the alloy should develop a n-type behavior. In the charge compensated alloy, CC, however the p-type behavior due to Mn substitution for Ti is compensated by the n-type behavior due to Sn substitution with Bi. The predicted magnetic behavior is based on the effective

VEC in the alloys which is expected to follow the Slater-Pauling prediction and half-Heusler alloys with VEC≠18 should exhibit a resultant magnetic behavior.(13, 30)

**Experimental Methods:**

The 5 different alloys were synthesized from 99.9 % pure elements in a vacuum arc melting system under a protective Ar gas environment. The ingots were flipped and melted several times to ensure complete mixing of the different elements. The ingots were further annealed at 800 $^0$C in vacuum for 120 hrs in order to homogenize the chemical composition and also to obtain a single phase. The annealed ingots were ground into powder and consolidated into circular discs of 12.7 mm diameter using the spark plasma sintering (SPS) process. The density in all the cases is found to be > 95 % of theoretical density. The phases present in each of the alloys and their structure, chemical composition were determined by X-ray diffraction and energy dispersive X-ray spectra (EDS) analysis in the scanning electron microscope respectively. The electrical resistivity in the temperature range 10 K to 1000 K was determined using the dc four probe technique. The low temperature resistivity and Seebeck coefficient were measured using a custom built system(31) while the high temperature properties were measured by a commercial system, Ulvac ZEM-2. Similarly the low temperature thermal conductivity was measured using a custom built system(32) while the high temperature thermal diffusivity was measured using NETZSCH LFA 457 system. The high temperature thermal conductivity is determined assuming the Dulong-Petit limit for the heat capacity. The magnetization as a function of both temperature and magnetic field were determined using the Quantum Design SQUID magnetometer.

**Results and Discussion:**

The microstructure of the different alloys as seen in scanning electron microscope, Figure 1, shows densely packed large grains with no porosity indicating bulk density after spark plasma sintering of the compacted alloy powders. The microstructure was found to be similar in all the alloys. The micrograph shown here is a SEM photo of microstructure in N2 alloy which exhibits more than one phase with the second phase in small quantity and has been identified in the x-ray diffraction pattern. The chemical composition determined using X-ray spectroscopy together with the designed chemical composition are given in Table 1. It can be seen that there is no significant loss of any of the elements in all the alloys except for Bi in the compensated alloy CC. In order to identify the phases present in different alloys and their crystal structures, Rietveld refinement of the X-ray diffraction patterns was done using the FULLPROF program.(33) The results of refinement of all the 5 different alloys together with the diffraction data are shown in Figure 2 and the refinement parameters are given in Table 2. The alloys P1 and P2 have a single cubic phase belonging to the space group F-43m while the N1 and N2 alloys have additional secondary phases in small fraction. The secondary phases are typically half-Heusler phases with slight variation in chemical composition or full-Heusler precipitates of extremely small size.(8, 34) The fraction of these secondary phases however does not exceed 20 % as all the major peaks in the diffraction pattern could be fitted to a single half-Heusler phase. The compensated alloy CC however has a single cubic phase confirming the formation of half-Heusler compound with no other phases. The position of 3 base elements Ti, Ni and Sn in the crystal structure in the 3 inequivalent positions 4a (0, 0, 0), 4b (1/2, 1/2, 1/2) and 4c (1/4, 1/4, 1/4) is critical for the electronic behavior of the alloys. Hence to determine the exact position of these elements in the synthesized alloys, the refinement was carried out by interchanging the position of Ti and Ni atoms in the structure. It was found that this did not alter the goodness of fit significantly in all the alloys. It was found that Sn occupies the 4b position while Ti and Ni occupy 4a and 4c positions in the crystal lattice respectively. This conforms to Type-III structure with an ionic bond between Ti/Ni and Sn and the other two bonds being covalent in nature. The structural stability determined by density functional theory calculations indicates that placing Ni atoms in the octahedral position, 4c, yields a structure with lowest energy compared to other possible configurations.(35-37) Our experimental results indeed confirm this observation and show that Ti and Sn form the NaCl type rock salt structure with the Ni atoms occupying the octahedral 4c sites. Hence the respective substituted atoms are expected to occupy the equivalent host element positions in the lattice.

The electrical resistivity of all the different alloys has been investigated as a function of temperature in the range 10 K to 1000 K and the results are shown in Figure 3 (a). The absolute resistivity in all the alloys is found to be low, varying between 0.06 to 5 mΩ cm at any temperature. The resistivity variation in the entire temperature range for all the alloys is small, i.e. the temperature coefficient of resistivity is extremely small. The temperature dependence of resistivity in the different category of alloys however is found to be different. In the case of p-type alloys, a metallic behavior with $\frac{d\rho}{dT} > 0$ is seen at low temperatures which changes to a semiconducting behavior with $\frac{d\rho}{dT} < 0$ at high temperatures. The resistivity exhibits a broad maximum in the range 300 K-500 K in the alloys P1 and P2. Such a behavior has been observed in rare-earth, R containing half-Heusler alloys, RPdBi/Sb which belong to the family of topological insulators exhibiting both superconductivity and magnetism.(38-43) The transition from semiconducting to metallic and subsequent superconducting behavior with decreasing temperatures in these rare-earth containing alloys has been found to be due to the band inversion phenomenon. In these alloys band inversion takes place due to topologically protected surface states when they are cooled to low temperatures. The variation of electrical resistivity with temperature in the non-superconducting state in the alloys P1 and P2 has been found to be due to a combination of both metallic and semiconducting channels operating simultaneously at any given temperature. The temperature dependence in such systems has been modeled using the combined equation given by;(44)

$$\rho(T) = \rho_m(T) + \rho_s(T) = (\rho_0^m + aT + bT^2) + (\rho_0^s \exp\left(-\frac{E_g}{2k_BT}\right)) \quad (1)$$

where $\rho_m$ and $\rho_s$ are metallic and semiconducting resistivities with $\rho_0^m$ and $\rho_0^s$ the residual resistivities, 'a' the electron-phonon scattering coefficient, 'b' the electron-electron scattering coefficient and 'Eg' semiconducting channel band gap. A fit of the electrical resistivity to equation (1) is shown in Figure 3 (a) and the different fit parameters are given in Table 3. The band gap 'Eg' for the semiconducting channel in the two alloys P1 and P2 is found to be 60 meV and 46.9 meV respectively, similar to the values determined in rare-earth containing half-Heusler alloys, RPdBi/Sb.(42-44) In the case of rare-earth containing half-Heusler alloys, the 'f' electrons of rare-earth element has been predicted to be the main reason for band inversion phenomenon leading to a superconducting behavior. In the present case however the alloys P1 and P2 do not contain any rare-earth elements and hence presence of defects such as anti-site substitutions and vacancies could lead to band inversion phenomenon.(45, 46)

The electrical resistivity of n-type alloys, N1, N2 as well as that of the compensated alloy CC increases with increasing temperature, $\frac{d\rho}{dT} > 0$ a typical metallic behavior. The resistivity of N1 increases continuously with temperature while that of N2 and CC indicate a saturation tendency at high temperatures. The electrical resistivity of these alloys can be modeled using the typical metallic behavior given by the first term in equation (1) and the results of fit are shown in Figure 3 and Table 3. It is interesting to note that the resistivity in N1, N2 and CC is essentially due to scattering of charge carriers by phonons with no significant contribution from electron-electron scattering. These alloys behave as typical degenerate semiconductors with phonon dominated carrier scattering.

The Seebeck coefficient 'α' measured as a function of temperature in the range 10 K to 1000 K of all the alloys is shown in Figure 4. The sign of 'α' in alloys P1, P2, N1 and N2 is in complete agreement with the alloy design considerations, i.e. P1 and P2 have a positive 'α' indicating holes as majority charge carriers while N1 and N2 have a negative 'α' with electrons as majority charge carriers. The Seebeck coefficient of CC however is extremely small and positive with extremely small variation with T. The Seebeck coefficient increases with increasing temperature in the 4 alloys, P1, P2, N1 and N2. The behavior of p-type alloys however is distinctly different compared to the n-type alloys. They exhibit a high temperature maximum between 600 K and 700 K, beyond which temperature bipolar transport becomes dominant and leads to a decrease of 'α'. These alloys also exhibit a clear phonon drag peak at low temperatures, 10 K to 20 K. It is observed in several other materials wherein the non-equilibrium distribution of phonons interacts with charge carriers and leads to 'dragging' phenomenon.(47-50) The alloy P1 however has an unusual temperature dependence of 'α'-an n-type behavior at low temperature which becomes p-type for T > 300K. This type of behavior in P2 however is not as pronounced compared to P1.The switching of the nature of charge carriers from electrons to holes at high temperature is a reflection of the charge transport behavior-electrical resistivity. The charge transport changes from being semiconducting at high temperature to a metallic behavior at low temperatures due to band inversion phenomenon and is accompanied by a change in the nature of charge carriers.

The Seebeck coefficient in general has two main contributions: One due to conventional thermal diffusion of charge carriers and second due to the non-equilibrium phonon distribution. The conventional thermal diffusion Seebeck coefficient, '$\alpha_d$' below the Debye temperature $\theta_D$ is given by the relation;(51)

$$\alpha_d = \left| -\left(\frac{\pi^2 k_B^2}{3eE_F}\right)T \right| \qquad (2)$$

where e is the carrier charge, $k_B$ the Boltzmann constant and '$E_F$' the Fermi energy. The Seebeck coefficient due to non-equilibrium phonon distribution on the other hand depends on the various scattering mechanisms and hence has a non-linear T-dependence. The linear variation of α of n-type alloys N1 and N2 indicates that thermal diffusion of charge carriers is dominant in these alloys and hence can be modeled using eq.(2). Hence α of these alloys has been fitted to eq.(2) to determine $E_F$. The effective charge carrier concentration n is then determined using the relation, n= $(2m^*E_F)/3\pi^2\hbar^3$ assuming an effective charge carrier mass of $2m_e$ for the charge carriers. The charge carrier concentration has been found to be $3.23\times10^{20}$ cm$^{-3}$ and $4.84\times10^{20}$ cm$^{-3}$ in the two alloys N1 and N2 respectively in agreement with the typical values found in half-Heusler alloys. The Seebeck coefficient of P1 and P2 alloys on the other hand has a highly non-linear behavior before the onset of bipolar conduction and is found to vary as $T^{3/2}$, a behavior observed in disordered systems.

The thermoelectric power factor described as $\alpha^2\sigma$ for the four alloys, P1, P2, N1 and N2 is shown in Figure 5(a). The n-type alloys in general have a higher power factor compared to the p-type alloys. The p-type alloy P2 has a maximum power factor of ~ 250 µWm$^{-1}$K$^{-2}$ at 700 K while N1 alloy has a power factor of ~ 900 µWm$^{-1}$K$^{-2}$ at 900 K. For thermoelectric application, both p- and n-type materials are essential and P2 and N1 couple can exhibit a high power factor at 700 K, a relatively high temperature for many applications. The main advantage of this couple is that they are made of identical base elements and hence will have compatible physical properties for potential application. The total thermal conductivity of the alloys in the range 10 K to 1000 K has been determined and is shown in Figure 5 (b). The p-type alloys have a much lower thermal conductivity compared to the n-type alloys at all temperatures.

The variation of magnetization M of the different alloys with both temperature T as well as external field H is shown in Figure 6. The variation of M in the different alloys is different and does not show similarities inspite of compositional similarities. The magnetization of p-type alloys increases with decreasing temperature at all fields, H, a behavior typical of paramagnetic materials. The susceptibility variation of P1 has been fitted to the Curie-Weiss law given by;

$$\chi(T) = \frac{N\mu_{eff}^2}{3k_B(T-T_C)} + \chi_0 \qquad (3)$$

where N is the effective number of magnetic atoms in the alloy with a magnetic moment µ$_{eff}$, $T_C$ the paramagnetic Curie temperature and $\chi_0$ the temperature independent susceptibility. The results are shown in Figure 6(a) along with experimental data. The temperature independent susceptibility $\chi_0$ is found to be 9.1×10-6 emu g$^{-1}$Oe$^{-1}$ and the Curie temperature $T_C$ is found to be -2 K. The near zero $T_C$ clearly indicates that the alloy behaves as a classical paramagnet without strong magnetic interactions between the Ni/Co atoms occupying the 4c site in the structure.(19, 44, 52, 53) The effective magnetic moment per atom is found to be 0.71 $\mu_B$, a value close to that of $Ni_{80}Co_{20}$ composition. The magnetic behavior of P2 alloy without 'Co' substitution is different compared to that of P1 alloy. The magnetization increases with decreasing T till~150 K with a curvature which is very different compared to a typical paramagnetic material. The magnetization increase is similar to that observed in ferromagnetic materials till this T. Below 150 K, the magnetization increase is similar to that observed in paramagnetic materials. These results show a magnetic transformation occurring at ~ 150 K, possibly due to rearrangement of anti-site defects at 4c. The temperature dependence of magnetization of n-type alloys which have Ni and Cu in the ratio 4:1 at 4c sites is markedly different when compared to the p-type alloys. The magnetization is nearly constant with an antiferromagnetic transition in the case of N1 alloy and a diamagnetic transition in the case of N2 alloy, transitions at ~ 20 K in both alloys. The temperature independence indicates that these alloys are Pauli paramagnets till the transition temperature and undergo a transition below 20 K into an antiferromagnetic/diamagnetic state respectively. The magnetization of CC alloy exhibits a typical ferromagnetic behavior with M increasing with decreasing T and increasing H. The temperature dependence has a typical ferromagnetic variation indicating that the presence of Mn results in altering the effective band structure and hence the magnetic behavior.

**Conclusions:**

The thermophysical and magnetic properties of 5 different half-Hesuler alloys based on Ti-Ni-Sn have been studied as a function of temperature. All the alloys exhibit thermophysical properties as per their designed chemical composition, i.e. p-type and n-type behavior at high temperatures. The low temperature properties however do not conform specifically to their valence electron count. The p-type alloys have a weak n-type character which transforms above room temperature. At extremely low temperatures, ~ 10 K a strong phonon drag is observed in these alloys which is not seen in the n-type alloys. They also exhibit a strong band inversion around room temperature, a phenomenon which has so far been observed only in rare-earth containing half-Heusler alloys. The absolute value of the Seebeck coefficient and total thermal conductivity in all the p-type and n-type alloys however are not ideal for thermoelectric application. Introduction of additional phonon scattering centers in the microstructure should result in lowering the overall thermal conductivity of the alloys. The Seebeck coefficient is relatively low when compared to the absolute resistivity of the alloys and this is not completely understood at present. The magnetic properties of the p-type and n-type alloys exhibits a paramagnetic behavior which can be understood based on their valence electron count. The temperature dependence of magnetization of n-type alloys however has a Pauli paramagnetic behavior, no temperature dependence, with a low temperature transition into an ordered magnetic state. These results clearly show that the electronic band structure of half-Heusler alloys is rather complex and can lead to a variety of electronic behaviors. A complete understanding of the electronic properties requires detailed band structure information.

**Acknowledgements:** The authors acknowledge Nanomission of Department of Science and Technology, Govt. of India for financial assistance and Prof. Terry Tritt for providing access to characterization facilities at Clemson University.

**References:**

1. Lin H, Wray LA, Xia Y, Xu S, Jia S, Cava RJ, et al. Half-Heusler ternary compounds as new multifunctional experimental platforms for topological quantum phenomena. Nature materials. 2010;9(7):546-9.
2. Chadov S, Qi X, Kubler J, Fecher GH, Felser C, Zhang SC. Tunable multifunctional topological insulators in ternary Heusler compounds. Nature materials. 2010;9(7):541-5.
3. Graf T, Felser C, Parkin SSP. Simple rules for the understanding of Heusler compounds. Prog Solid State Ch. 2011;39(1):1-50.
4. Felser C, Wollmann L, Chadov S, Fecher GH, Parkin SSP. Basics and prospective of magnetic Heusler compounds. Apl Mater. 2015;3(4).
5. Dederichs PH, Galanakis I, Mavropoulos P. Half-metallic alloys: electronic structure, magnetism and spin polarization. J Electron Microsc. 2005;54:I52-I5.
6. Xiong DB, Okamoto NL, Waki T, Zhao YF, Kishida K, Inui H. High-TC Ferromagnetic Semiconductor-Like Behavior and Unusual Electrical Properties in Compounds with a 222 Superstructure of the Half-Heusler Phase. Chem-Eur J. 2012;18(9):2536-42.
7. Chai YW, Oniki T, Kimura Y. Microstructure and thermoelectric properties of a ZrNi1.1Sn half-Heusler alloy. Acta Mater. 2015;85:290-300.
8. Downie RA, MacLaren DA, Bos JWG. Thermoelectric performance of multiphase XNiSn (X = Ti, Zr, Hf) half-Heusler alloys. J Mater Chem A. 2014;2(17):6107-14.
9. Poon SJ, Wu D, Zhu S, Xie WJ, Tritt TM, Thomas P, et al. Half-Heusler phases and nanocomposites as emerging high-ZT thermoelectric materials. J Mater Res. 2011;26(22):2795-802.
10. Yang J, Li HM, Wu T, Zhang WQ, Chen LD, Yang JH. Evaluation of Half-Heusler Compounds as Thermoelectric Materials Based on the Calculated Electrical Transport Properties. Adv Funct Mater. 2008;18(19):2880-8.
11. Fecher GH, Rausch E, Balke B, Weidenkaff A, Felser C. Half-Heusler materials as model systems for phase-separated thermoelectrics. Phys Status Solidi A. 2016;213(3):716-31.

12. Bos JWG, Downie RA. Half-Heusler thermoelectrics: a complex class of materials. J Phys-Condens Mat. 2014;26(43).
13. Galanakis I, Dederichs PH, Papanikolaou N. Origin and properties of the gap in the half-ferromagnetic Heusler alloys. Phys Rev B. 2002;66(13).
14. Joshi G, Dahal T, Chen S, Wang HZ, Shiomi J, Chen G, et al. Enhancement of thermoelectric figure-of-merit at low temperatures by titanium substitution for hafnium in n-type half-Heuslers Hf0.75-xTixZr0.25NiSn0.99Sb0.01. Nano Energy. 2013;2(1):82-7.
15. Balke B, Ouardi S, Graf T, Barth J, Blum CGF, Fecher GH, et al. Seebeck coefficients of half-metallic ferromagnets. Solid State Commun. 2010;150(11-12):529-32.
16. Galanakis I, Ozdogan K, Sasioglu E, Aktas B. Ferrimagnetism and antiferromagnetism in half-metallic Heusler alloys. Phys Status Solidi A. 2008;205(5):1036-9.
17. Fukushima T, Sato K, Katayama-Yoshida H, Dederichs PH. Dilute magnetic semiconductors based on half-Heusler alloys. J Phys Soc Jpn. 2007;76(9).
18. Otto MJ, Vanwoerden RAM, Vandervalk PJ, Wijngaard J, Vanbruggen CF, Haas C, et al. Half-Metallic Ferromagnets .1. Structure and Magnetic-Properties of Nimnsb and Related Inter-Metallic Compounds. J Phys-Condens Mat. 1989;1(13):2341-50.
19. Tobola J, Pierre J, Kaprzyk S, Skolozdra RV, Kouacou MA. Crossover from semiconductor to magnetic metal in semi-Heusler phases as a function of valence electron concentration. J Phys-Condens Mat. 1998;10(5):1013-32.
20. Wollmann L, Chadov S, Kubler J, Felser C. Magnetism in cubic manganese-rich Heusler compounds. Phys Rev B. 2014;90(21).
21. Ouardi S, Fecher GH, Balke B, Kozina X, Stryganyuk G, Felser C, et al. Electronic transport properties of electron- and hole-doped semiconducting C1b Heusler compounds: NiTi1-xMxSn (M=Sc, V). Phys Rev B. 2010;82(8).
22. Hohl H, Ramirez AP, Goldmann C, Ernst G, Wolfing B, Bucher E. Efficient dopants for ZrNiSn-based thermoelectric materials. J Phys-Condens Mat. 1999;11(7):1697-709.
23. Uher C, Yang J, Hu S, Morelli DT, Meisner GP. Transport properties of pure and doped MNiSn (M=Zr, Hf). Phys Rev B. 1999;59(13):8615-21.
24. Simonson JW, Wu D, Xie WJ, Tritt TM, Poon SJ. Introduction of resonant states and enhancement of thermoelectric properties in half-Heusler alloys. Phys Rev B. 2011;83(23).
25. Chen L, Gao S, Zeng X, Dehkordi AM, Tritt TM, Poon SJ. Uncovering high thermoelectric figure of merit in (Hf,Zr)NiSn half-Heusler alloys. Appl Phys Lett. 2015;107(4).
26. Bartholome K, Balke B, Zuckermann D, Kohne M, Muller M, Tarantik K, et al. Thermoelectric Modules Based on Half-Heusler Materials Produced in Large Quantities. J Electron Mater. 2014;43(6):1775-81.
27. Culp SR, Poon SJ, Hickman N, Tritt TM, Blumm J. Effect of substitutions on the thermoelectric figure of merit of half-Heusler phases at 800 degrees C. Appl Phys Lett. 2006;88(4).
28. Downie RA, MacLaren DA, Smith RI, Bos JWG. Enhanced thermoelectric performance in TiNiSn-based half-Heuslers. Chem Commun. 2013;49(39):4184-6.
29. Downie RA, Popuri SR, Ning HP, Reece MJ, Bos JWG. Effect of Spark Plasma Sintering on the Structure and Properties of Ti1-xZrxNiSn Half-Heusler Alloys. Materials. 2014;7(10):7093-104.
30. Galanakis I, Dederichs PH, Papanikolaou N. Slater-Pauling behavior and origin of the half-metallicity of the full-Heusler alloys. Phys Rev B. 2002;66(17).
31. Pope AL, Littleton RT, Tritt TM. Apparatus for the rapid measurement of electrical transport properties for both "needle-like" and bulk materials. Rev Sci Instrum. 2001;72(7):3129-31.
32. Pope AL, Zawilski B, Tritt TM. Description of removable sample mount apparatus for rapid thermal conductivity measurements. Cryogenics. 2001;41(10):725-31.
33. Rodriguez-Caravajal J. Physics B. 1993;192:55-65.

34. Chai YW, Kimura Y. Microstructure evolution of nanoprecipitates in half-Heusler TiNiSn alloys. Acta Mater. 2013;61(18):6684-97.
35. Nanda BRK, Dasgupta I. Electronic structure and magnetism in half-Heusler compounds. J Phys-Condens Mat. 2003;15(43):7307-23.
36. Larson P, Mahanti SD, Kanatzidis MG. Structural stability of Ni-containing half-Heusler compounds. Phys Rev B. 2000;62(19):12754-62.
37. Ogut S, Rabe KM. Band-Gap and Stability in the Ternary Intermetallic Compounds Nisnm (M=Ti,Zr,Hf) - a First-Principles Study. Phys Rev B. 1995;51(16):10443-53.
38. Nakajima Y, Hu R, Kirshenbaum K, Hughes A, Syers P, Wang X, et al. Topological RPdBi half-Heusler semimetals: A new family of noncentrosymmetric magnetic superconductors. Science advances. 2015;1(5):e1500242.
39. Nikitin AM, Pan Y, Mao X, Jehee R, Araizi GK, Huang YK, et al. Magnetic and superconducting phase diagram of the half-Heusler topological semimetal HoPdBi. Journal of physics Condensed matter : an Institute of Physics journal. 2015;27(27):275701.
40. Meinert M. Unconventional Superconductivity in YPtBi and Related Topological Semimetals. Physical review letters. 2016;116(13):137001.
41. Pavlosiuk O, Kaczorowski D, Wisniewski P. Shubnikov - de Haas oscillations, weak antilocalization effect and large linear magnetoresistance in the putative topological superconductor LuPdBi. Sci Rep-Uk. 2015;5.
42. Gofryk K, Kaczorowski D, Plackowski T, Leithe-Jasper A, Grin Y. Magnetic and transport properties of rare-earth-based half-Heusler phases RPdBi: Prospective systems for topological quantum phenomena. Phys Rev B. 2011;84(3).
43. Gofryk K, Kaczorowski D, Plackowski T, Leithe-Jasper A, Grin Y. Magnetic and transport properties of the rare-earth-based Heusler phases RPdZ and RPd(2)Z (Z=Sb,Bi). Phys Rev B. 2005;72(9).
44. Pavlosiuk O, Kaczorowski D, Fabreges X, Gukasov A, Wisniewski P. Antiferromagnetism and superconductivity in the half-Heusler semimetal HoPdBi. Sci Rep-Uk. 2016;6.
45. Xie HH, Wang H, Fu CG, Liu YT, Snyder GJ, Zhao XB, et al. The intrinsic disorder related alloy scattering in ZrNiSn half-Heusler thermoelectric materials. Sci Rep-Uk. 2014;4.
46. Qiu PF, Yang J, Huang XY, Chen XH, Chen LD. Effect of antisite defects on band structure and thermoelectric performance of ZrNiSn half-Heusler alloys. Appl Phys Lett. 2010;96(15).
47. Pokharel M, Zhao HZ, Lukas K, Ren ZF, Opeil C, Mihaila B. Phonon drag effect in nanocomposite FeSb2. Mrs Commun. 2013;3(1):31-6.
48. Zlatic V, Boyd GR, Freericks JK. Universal thermopower of bad metals. Phys Rev B. 2014;89(15).
49. Belashchenkov KD, Livanov DV. Effect of impurities on thermoelectric power due to phonon drag in metals. J Exp Theor Phys+. 1997;84(6):1221-4.
50. Boxus J, Issi JP. Giant Negative Phonon Drag Thermopower in Pure Bismuth. J Phys C Solid State. 1977;10(15):L397-L401.
51. Bernard RD. London: Taylor and Francis Ltd. 1972.
52. Hartjes K, Jeitschko W. Crystal-Structures and Magnetic-Properties of the Lanthanoid Nickel Antimonides Lnnisb (Ln=La-Nd, Sm, Gd-Tm, Lu). J Alloy Compd. 1995;226(1-2):81-6.
53. Damewood L, Busemeyer B, Shaughnessy M, Fong CY, Yang LH, Felser C. Stabilizing and increasing the magnetic moment of half-metals: The role of Li in half-Heusler LiMnZ (Z=N, P, Si). Phys Rev B. 2015;91(6).

**Table 1:** Designed alloy chemical composition along with predicted transport and magnetic behavior of all HH alloys.

| Sl. No. | Designed Chemical Composition | Actual Chemical Composition | Type of Charge carrier | VEC | Magnetic Moment, $\mu B$ |
|---|---|---|---|---|---|
| 1. | $(Ti_{0.5}Zr_{0.5})(Ni_{0.8}Co_{0.2})Sn$ P1 | $(Ti_{0.47}Zr_{0.52})(Ni_{0.84}Co_{0.21})Sn_{0.95}$ | p-type | 18 | 0 |
| 2. | $(Ti_{0.5}Zr_{0.25}Sc_{0.25})Ni(Sn_{0.95}In_{0.05})$ P2 | $(Ti_{0.44}Zr_{0.22}Sc_{0.29})Ni(SnIn_{0.032})$ | p-type | 18 | 0 |
| 3. | $(Ti_{0.8}Nb_{0.2})(Ni_{0.8}Cu_{0.2})(Sn_{0.95}Sb_{0.05})$ N1 | $(Ti_{0.71}Nb_{0.13})(Ni_{0.8}Cu_{0.2})(Sn_{1.09}Sb_{0.059})$ | n-type | 19 | 1 |
| 4. | $(Ti_{0.5}Nb_{0.5})(Ni_{0.8}Cu_{0.2})(Sn_{0.95}Sb_{0.05})$ N2 | $(Ti_{0.5}Nb_{0.3})(Ni_{0.8}Cu_{0.18})(Sn_{1.18}Sb_{0.06})$ | n-type | 19 | 1 |
| 5. | $(Ti_{0.2}Mn_{0.8})Ni(Sn_{0.9}Bi_{0.1})$ CC | $(Ti_{0.21}Mn_{0.8})Ni_{1.09}(Sn_{0.9}Bi_{0.004})$ | Compensated | 20 | 2 |

**Table 2:** Lattice parameter and atomic positions in all the half-Heusler compounds obtained by Rietveld refinement. All the alloys have a cubic F-43m crystal structure.

| Sl. No. | Designed Chemical Composition | Lattice Parameter, A | 4a | 4b | 4c |
|---|---|---|---|---|---|
| 1. | $(Ti_{0.5}Zr_{0.5})(Ni_{0.8}Co_{0.2})Sn$ P1 | 6.012 | Ni, Co | Sn | Ti, Zr |
| 2. | $(Ti_{0.5}Zr_{0.25}Sc_{0.25})Ni(Sn_{0.95}In_{0.05})$ P2 | 6.075 | Ni | Sn, In | Ti, Zr, Sc |
| 3. | $(Ti_{0.8}Nb_{0.2})(Ni_{0.8}Cu_{0.2})(Sn_{0.95}Sb_{0.05})$ N1 | 6.013 | Ni, Cu | Sn, Sb | Ti, Nb |
| 4. | $(Ti_{0.5}Nb_{0.5})(Ni_{0.8}Cu_{0.2})(Sn_{0.95}Sb_{0.05})$ N2 | 6.011 | Ni, Cu | Sn, Sb | Ti, Nb |
| 5. | $(Ti_{0.2}Mn_{0.8})Ni(Sn_{0.9}Bi_{0.1})$ CC | 6.034 | Ni | Sn, Bi | Ti, Mn |

**Table 3:** The electronic transport parameters of the different half-Heusler alloys is given here. These parameters are obtained by fitting the resistivity to eq. (1).

| Compounds | $\rho_0^m$, mΩ cm | a, mΩ cm K$^{-1}$ | b, mΩ cm K$^{-2}$ | $\rho_0^s$, mΩ cm | Eg, meV |
|---|---|---|---|---|---|
| P1 | 2.36 | 2.06 x 10$^{-4}$ | -1.70 x 10$^{-5}$ | 7.66 | 60 |
| P2 | 0.85 | 9.26 x 10$^{-4}$ | -9.36 x 10$^{-7}$ | 0.25 | 46.9 |
| N1 | 0.071 | 1.16 x 10$^{-4}$ | 1.94 x 10$^{-8}$ | - | - |
| N2 | 0.059 | 1.39 x 10$^{-4}$ | -7.06 x 10$^{-8}$ | - | - |
| CC | 0.17 | 1.04 x 10$^{-4}$ | -9.51 x 10$^{-9}$ | - | - |

**Figure Captions:**

**Figure 1.** The scanning electron micrograph shows a dense polycrystalline microstructure with large grains after spark plasma sintering. The microstructure shown here is from N2 alloy which has small amount of secondary phases along with the half-Heusler phase.

**Figure 2.** The x-ray diffraction patterns obtained from the 5 alloys shows the presence of half-Heusler phase in all the cases. In the n-type alloys however secondary phases are also seen in small quantity apart from the half-Heusler phase.

**Figure 3.** The electrical resistivity variation with temperature of the alloys exhibits a contrasting behavior. The p-type alloys undergo a metal to semiconducting transition on heating, (a) while the n-type and compensated alloys exhibit a degenerate semiconducting behavior, (b) with a weak positive temperature dependence. The line through the data points is a fit to eq. (1).

**Figure 4.** The variation of Seebeck coefficient α with temperature shows that alloys P1 and P2 have holes as predominant charge carriers for T > 300 K while in the N1 and N2 alloys electrons are the predominant charge carriers at all temperatures. The temperature dependence in the two cases is also very different. The low temperature variation of α shown in the inset illustrates the effect of phonon drag in p-type alloys.

**Figure 5.** The power factor variation with temperature of the different alloys is shown in (a) while the variation of thermal conductivity is shown in (b). The thermal conductivity of the alloys is high for thermoelectric application.

**Figure 6.** The variation of magnetization M with temperature T and external magnetic field H is shown. The p-type and n-type alloys are paramagnetic while the compensated alloy CC is ferromagnetic in nature.

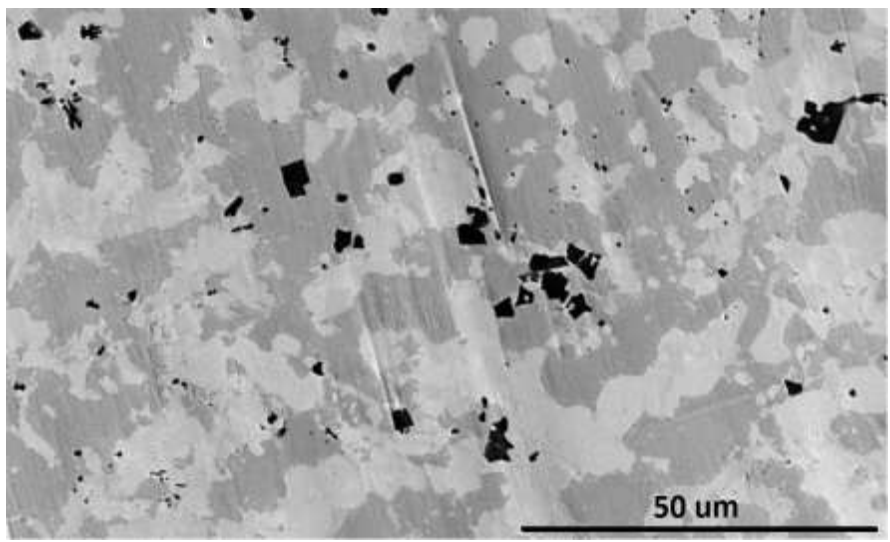

**Figure 1**

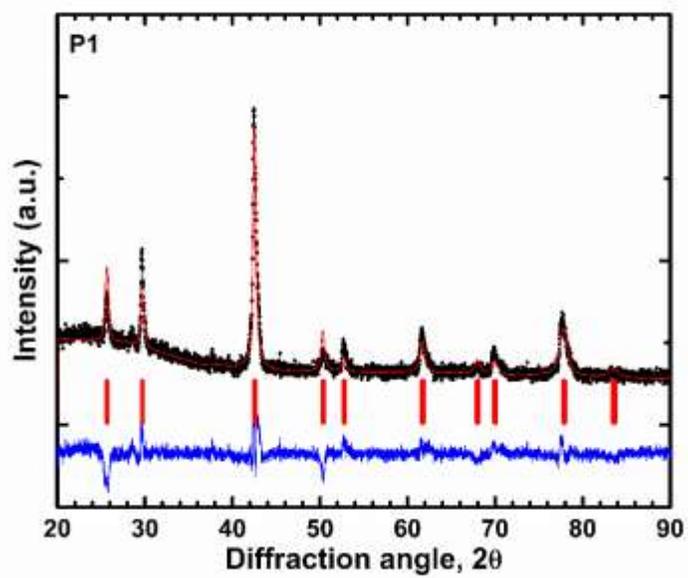

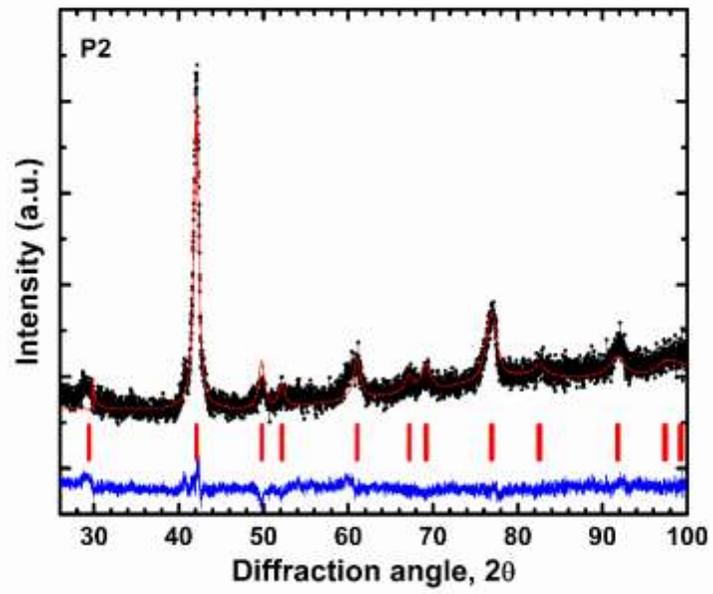

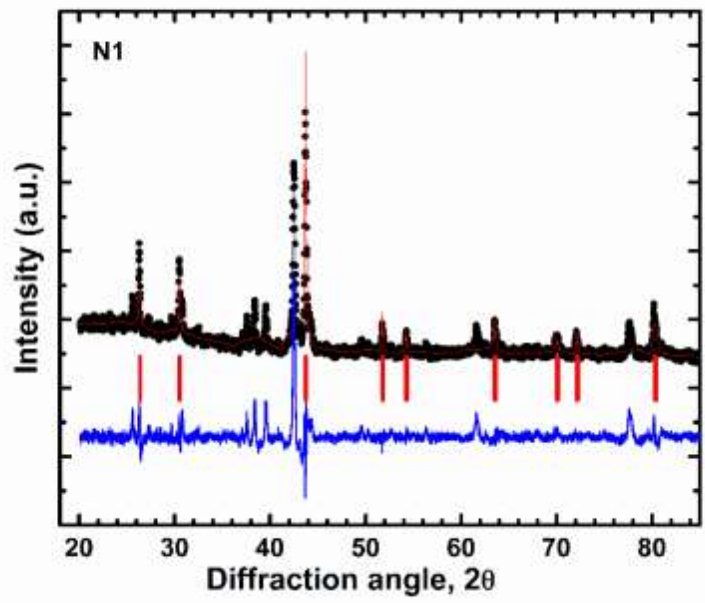

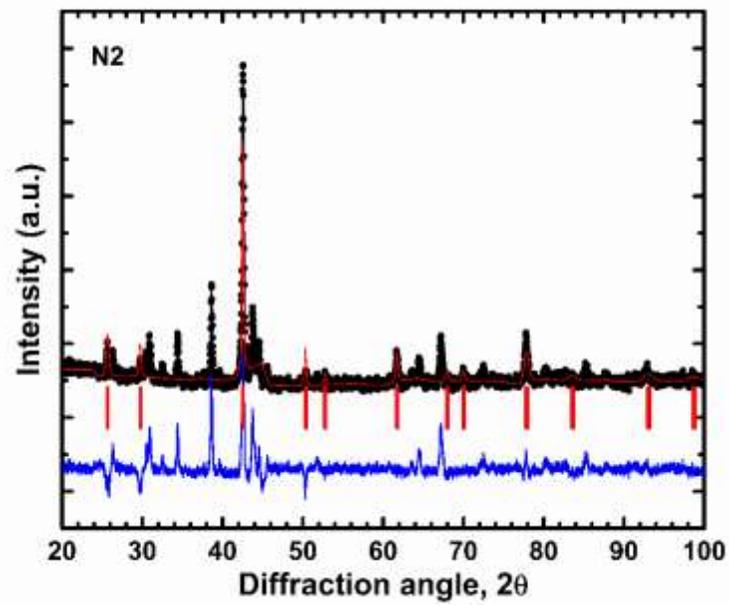

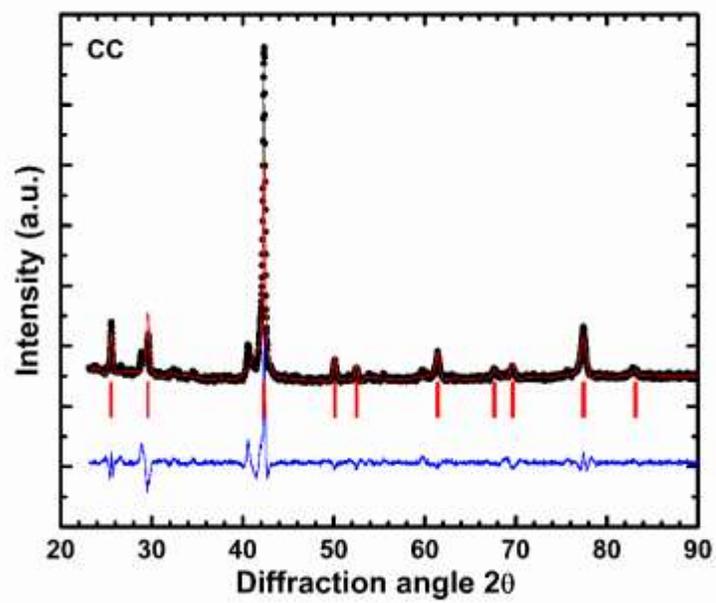

**Figure 2**

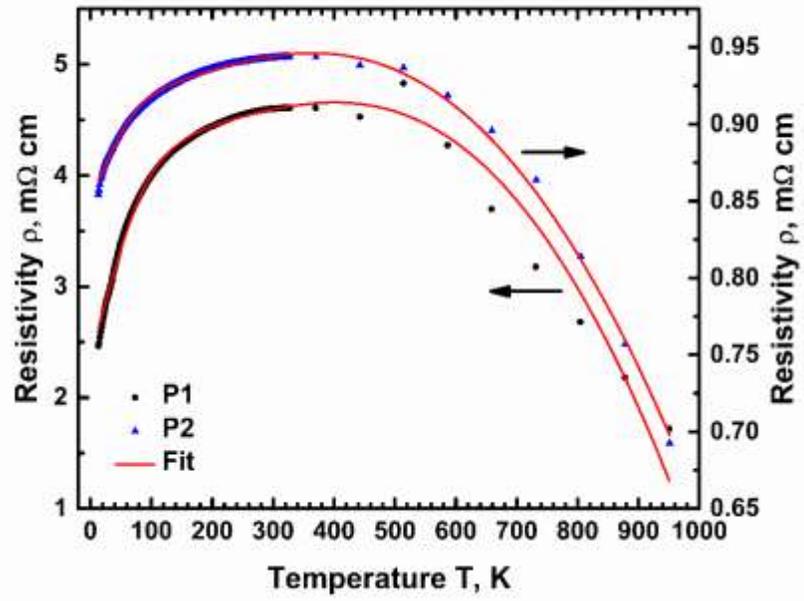

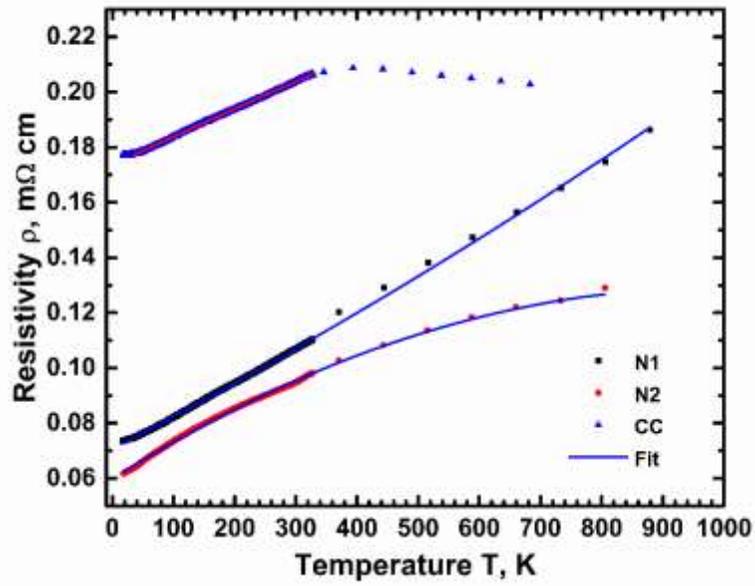

**Figure 3**

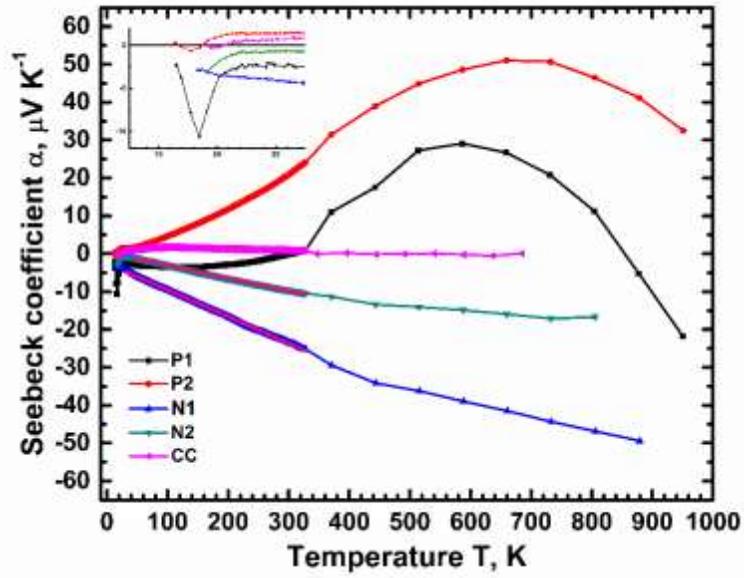

**Figure 4**

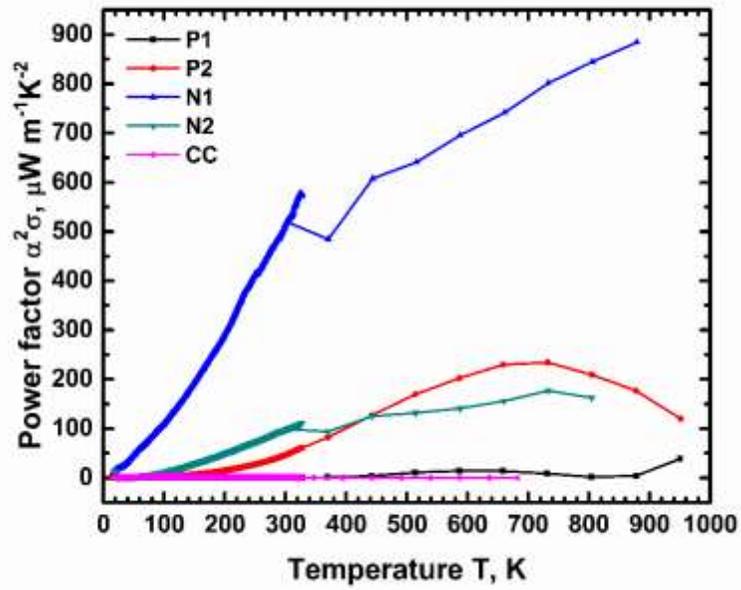

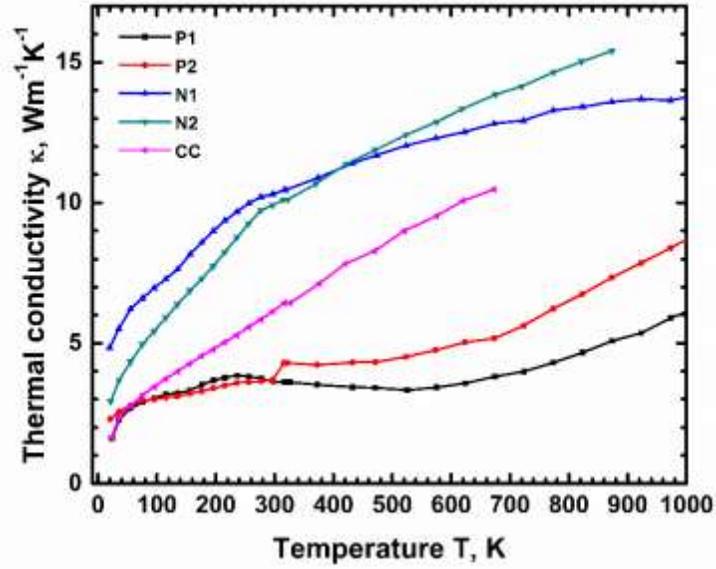

**Figure 5**

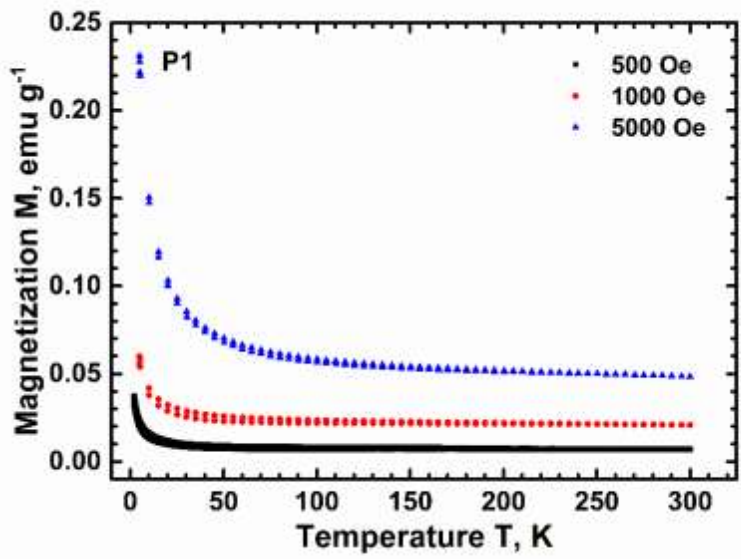

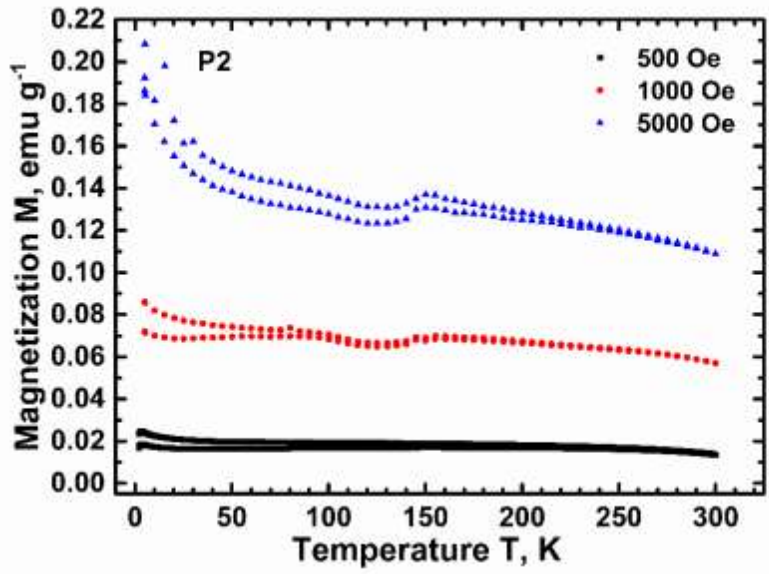

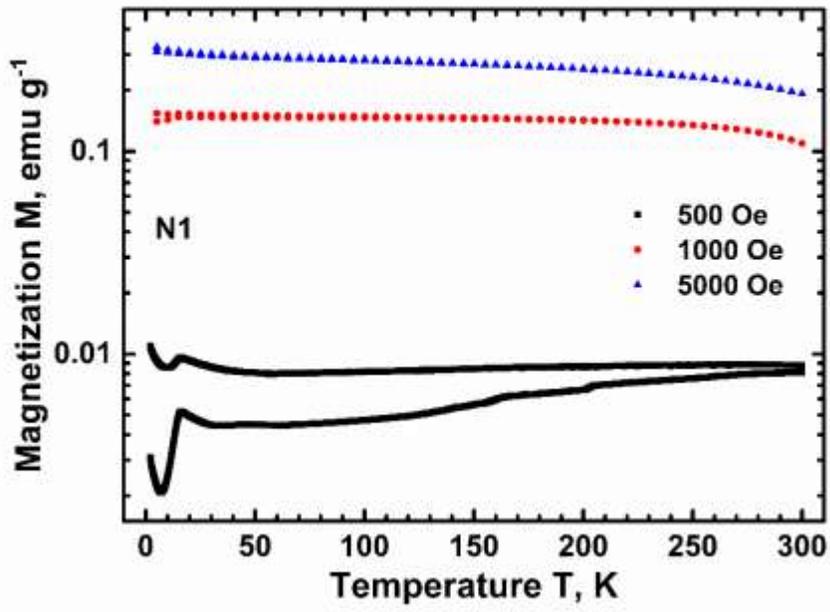

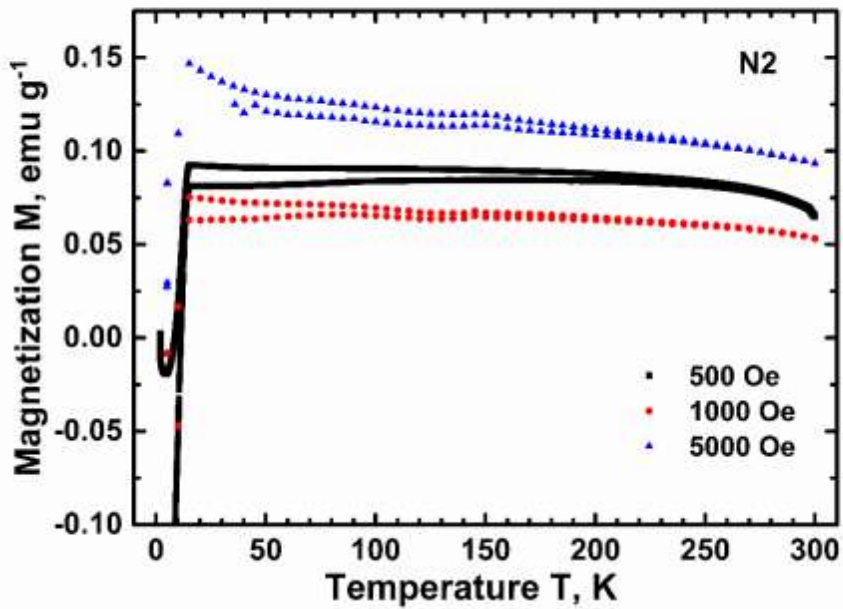

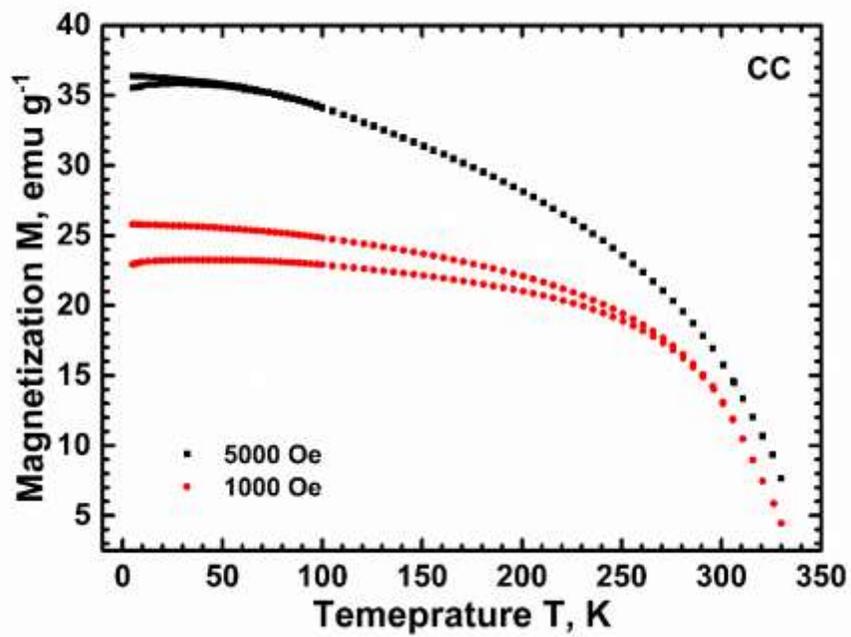

**Figure 6**